\documentclass[conference]{IEEEtran}
\IEEEoverridecommandlockouts
\usepackage{cite}
\usepackage{amsmath,amssymb,amsfonts}
\usepackage{algorithmic}
\usepackage{graphicx}
\usepackage{textcomp}
\usepackage{xcolor}
\usepackage{array}
\usepackage{balance}
\def\BibTeX{{\rm B\kern-.05em{\sc i\kern-.025em b}\kern-.08em
    T\kern-.1667em\lower.7ex\hbox{E}\kern-.125emX}}
\begin{document}

\title{Revolutionizing Undergraduate Learning: CourseGPT and Its Generative AI Advancements\\
%{\footnotesize \textsuperscript{*}Note: Sub-titles are not captured in Xplore and should not be used}
%\thanks{Identify applicable funding agency here. If none, delete this.}
}
\author{\IEEEauthorblockN{Ahmad M. Nazar}
\IEEEauthorblockA{\textit{Dept. of Elec. \& Comp. Eng.} \\
\textit{Iowa State University}\\
Ames, IA, USA \\
amnazar@iastate.edu}
\and
\IEEEauthorblockN{Mohamed Y. Selim}
\IEEEauthorblockA{\textit{Dept. of Elec. \& Comp. Eng.} \\
\textit{Iowa State University}\\
Ames, IA, USA \\
myoussef@iastate.edu}
\and
\IEEEauthorblockN{Ashraf Gaffar}
\IEEEauthorblockA{\textit{Dept. of Elec. \& Comp. Eng.} \\
\textit{Iowa State University}\\
Ames, IA, USA \\
gaffar@iastate.edu}
\and
\IEEEauthorblockN{Shakil Ahmed}
\IEEEauthorblockA{\textit{Dept. of Elec. \& Comp. Eng.} \\
\textit{Iowa State University}\\
Ames, IA, USA \\
shakil@iastate.edu}
}

\maketitle

\begin{abstract}
Integrating Generative AI (GenAI) into educational contexts presents a transformative potential for enhancing learning experiences. This paper introduces CourseGPT, a generative AI tool designed to support instructors and enhance the educational experiences of undergraduate students. Built on open-source Large Language Models (LLMs) from Mistral AI, CourseGPT offers continuous instructor support and regular updates to course materials, enriching the learning environment. By utilizing course-specific content, such as slide decks and supplementary readings and references, CourseGPT provides precise, dynamically generated responses to student inquiries. Unlike generic AI models, CourseGPT allows instructors to manage and control the responses, thus extending the course scope without overwhelming details.
The paper demonstrates the application of CourseGPT using the CPR E 431: Basics of Information System Security course as a pilot. This course, with its large enrollment and diverse curriculum, serves as an ideal testbed for CourseGPT. The tool aims to enhance the learning experience, accelerate feedback processes, and streamline administrative tasks. The study evaluates CourseGPT’s impact on student outcomes, focusing on correctness scores, context recall, and faithfulness of responses. Results indicate that the Mixtral-8x7b model, with a higher parameter count, outperforms smaller models, achieving an 88.0\% correctness score and a 66.6\% faithfulness score.
Additionally, feedback from former students and teaching assistants on CourseGPT's accuracy, helpfulness, and overall performance was collected. The outcomes revealed that a significant majority found CourseGPT to be highly accurate and beneficial in addressing their queries, with many praising its ability to provide timely and relevant information.
%While CourseGPT shows promising advancements, challenges such as ethical implications, data privacy, and the need for ongoing instructor training and support remain. CourseGPT holds the potential to revolutionize teaching and learning practices, benefiting students, instructors, and educational institutions.
\end{abstract}


\section{Introduction}
There is a growing recognition of the transformative potential of AI-driven technologies in enhancing academic and student learning experiences as the educational landscape continues to evolve. Integrating intelligent virtual assistants powered by LLMs is a promising solution for improving student interactions with course materials and seeking academic guidance. Adopting LLM-based assistants in educational settings represents a shift towards personalized and interactive learning environments with domain-specific knowledge. These assistants provide personalized support, real-time feedback, and domain-specific knowledge access, enabling deeper student engagement and more effective learning outcomes.

LLMs are known to have broad knowledge on various corpus of data, unless they have been fine-tuned. Fine-tuning LLMS is an exhaustive and time-consuming process. A method substituting fine-tuning methods is Retrieval Augmented Generation (RAG) which provides a knowledge base for the LLM to utilize in answering prompts \cite{ragSurvey}. This work evaluates RAG-based-LLM effectiveness as intelligent university course assistants. We aim to assess how different LLM sizes, ranging from 7 billion to 47 billion parameters, impact their ability to assist students with questions regarding the course. We identify the strengths and limitations of each LLM by establishing a shared knowledge base and employing rigorous evaluation methodologies. LLM assessment includes analyzing the correctness, context recall, and faithfulness of responses generated by each LLM within the course's scope. CourseGPT utilizes Mistral LLMs \cite{mistral, mixtral}.

CourseGPT is fine-tuned for the Iowa State University CPR E 431: Introduction to Wireless Network Security course. CourseGPT was deployed on a server and provided to a few students who completed the course in previous semesters and a previous teaching assistant. These participants were surveyed on CourseGPT's accuracy, helpfulness, and performance. 

An LLM assistant can be integrated into a virtual teaching assistant using GPT3. The virtual intelligent teaching assistant system at the framework's core is a voice-enabled helper capable of answering various course-specific questions, such as administrative and logistical questions and course policies \cite{another-ai-ta}. Integrating an RAG-LLM-based assistant within a university system marks a notable advancement in student support mechanisms \cite{ai-ta}. The AI-teaching assistant operates like traditional university forums while attending to student inquiries with interactive and informative capabilities. 

It is worth noting that the recent CourseGPT-zh is independent of our CourseGPT. CourseGPT-zh is a course-oriented education LLM that leverages a high-quality question-answering corpus distillation framework with prompt optimization. CourseGPT-zh introduces a novel method for discrete prompt optimization based on LLM-as-Judge, aligning LLM responses with user needs while saving response length \cite{other-course-gpt}. Our CourseGPT involves domain-specific knowledge to deploy as a student assistant that provides answers to questions to specific class-related knowledge. 

This work is organized as follows: Section \ref{system model} shows CourseGPT's design and implementation technical details. Section \ref{methodology} describes RAG-LLM evaluation methodologies and metrics. RAG-LLM evaluation results and analyses achieved from deploying CourseGPT are described in Section \ref{results}. The challenges associated with RAG-LLM deployment and fine-tuning are discussed in Section \ref{challenges}, and Section \ref{conclusion} concludes our work on LLM advancements in education.

\section{CourseGPT: Improving Student Success}
\label{Impact}
\subsection{CourseGPT Impact}

Undergraduate teaching often necessitates concise instructions within fine-grained modules and more individualized support to guide students step by step, ideally on a weekly or bi-weekly basis. CourseGPT enhances this approach by enabling instructors to create detailed, modular content using its interactive interface. This tool provides students with personalized support through immediate answers and real-time guidance, encouraging them to ask more questions without fear of judgment. Consequently, students can focus on and complete each module in a single interactive session, significantly reducing the frustration associated with waiting for responses from instructors or teaching assistants.

The immediate and personalized feedback provided by CourseGPT promotes an excellent educational experience by increasing student engagement and productivity. Students are more inclined to seek help and clarification on various topics, knowing they will receive prompt and relevant responses. This not only enhances their understanding of the course material but also builds their confidence in mastering complex subjects. By fostering a more interactive and responsive learning environment, CourseGPT aligns with any school's strategic goals of promoting educational experience excellence and enhancing knowledge and discovery.

CourseGPT’s versatility makes it suitable for a wide range of courses with different instructional materials, including mathematics, programming, and natural language artifacts. The success of similar AI tools, like ChatGPT, across various domains evidences this adaptability. Our pilot class, CPR E 431: Basics of Information System Security, includes all three types of materials, demonstrating CourseGPT's capability to handle diverse content. This pilot will serve as a proof of concept that can be generalized to a broader range of courses, benefiting a larger number of students across the Arts, Sciences, and Engineering disciplines. Although the initial pilot will not address materials such as drawings and graphics, future iterations of CourseGPT may expand to include these as well, further broadening its applicability.

\subsection{Enhancing Students' Learning Outcomes}
CourseGPT is designed to be a valuable tool for both instructors and students by providing guided procedures to implement three specific student outcomes from ABET: SO1, SO4, and SO7. This tool aims to enhance learning outcomes and also offers structured guidance for instructors, improving the overall student learning experience, whether or not the class seeks ABET accreditation.

\subsubsection{SO1: An Ability to Identify, Formulate, and Solve Complex Engineering Problems}
CourseGPT assists instructors in structuring their material into three phases: identification, formulation, and solution of problems. Initially, students engage in broad, targeted searches and ask multiple questions to identify problems. Next, they receive detailed feedback to help formulate these problems accurately. Finally, students are guided to solve these problems by asking further questions and exploring similar solutions to validate their approaches. CourseGPT's interactive, real-time feedback and individualized support during these sessions enhance students' problem-solving skills and deepen their understanding of complex engineering challenges.

\subsubsection{SO4: An Ability to Recognize Ethical and Professional Responsibilities}
Leveraging years of teaching experience, CourseGPT significantly enhances students' ability to recognize ethical and professional responsibilities. It promotes critical thinking by linking a project's impact to global, economic, environmental, and societal contexts. Through long discussions and repeated, focused questions, CourseGPT facilitates intelligent, real-time interactive sessions that deepen students' understanding of the broader implications of their projects. This structured approach helps students develop a more comprehensive perspective on their ethical and professional responsibilities.

\subsubsection{SO7: An Ability to Acquire and Apply New Knowledge}
Achieving this outcome requires going beyond traditional classroom instruction. CourseGPT addresses this need by allowing instructors to compile supplemental material and facilitate interactive discussions between students and the tool. This capability helps students navigate the vast amount of information available on the internet and apply new knowledge effectively. CourseGPT's precise guidance and real-time support enable students to acquire and integrate new knowledge seamlessly into their existing understanding.

In summary, CourseGPT represents a significant advancement in educational technology, providing personalized, real-time support that enhances undergraduate learning experiences and outcomes. By facilitating complex problem-solving, promoting ethical responsibilities, and encouraging the acquisition of new knowledge, CourseGPT aligns with strategic educational goals. The pilot phase will establish a foundation for broader implementation, benefiting a larger cohort of students and setting a new standard for educational excellence in higher education.

\section{CourseGPT Implementation}
\label{system model}
CourseGPT expends advanced LLM computational techniques with RAGs at the framework's core. RAGs enable knowledge base integration into the generative process. This integration is vital in ensuring generated content relevance and accuracy, especially in a university course environment where new findings and teaching material are added. We chose text embeddings that satisfy our operational constraints to construct a robust knowledge base system while optimizing retrieval precision. This knowledge base guarantees that the retrieved information is relevant and contextually appropriate for the generative tasks. CourseGPT relies on LLMs as they transform raw RAG data and text embeddings into coherent and contextually rich responses to user prompts. 

\begin{figure}
    \centering
    \includegraphics[scale = 0.4]{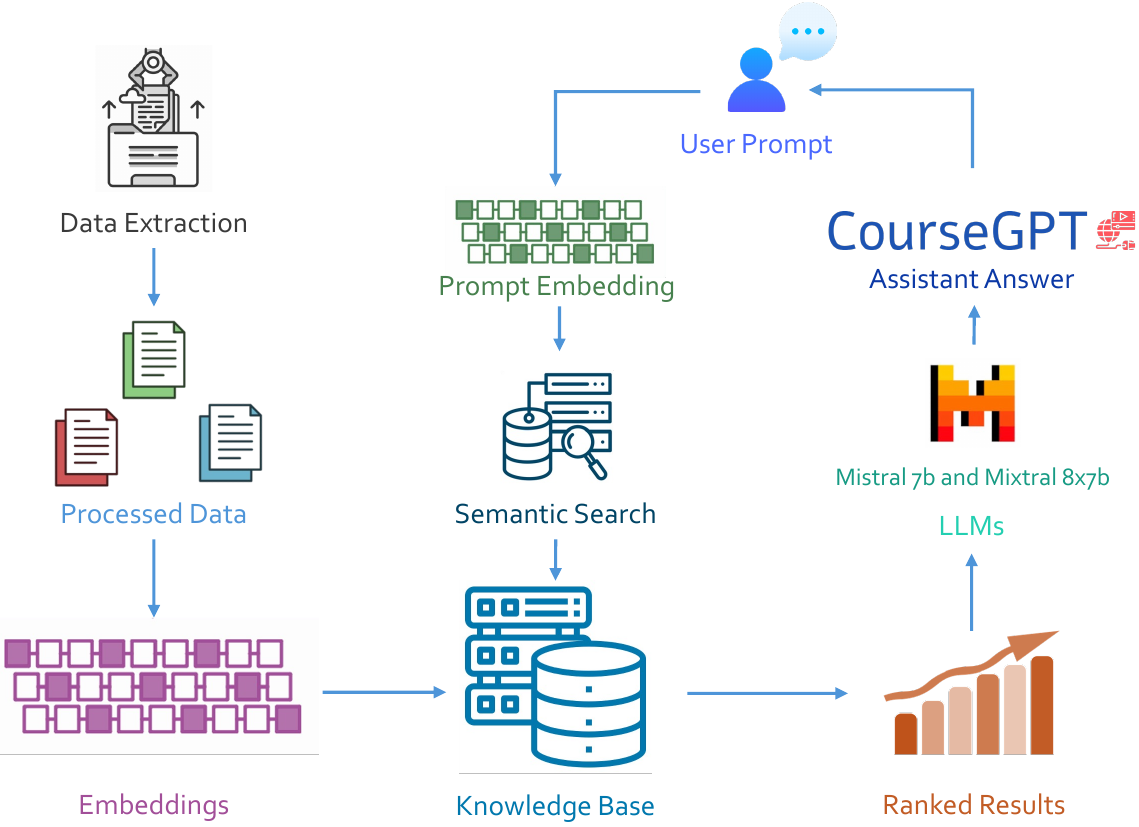}
    \caption{CourseGPT Workflow}
    \label{fig:course-gpt-workflow}
\end{figure}

This section provides details of each CourseGPT component, clarifying their contributions and collective impact on the system's functionality. By exploring these elements, we offer insights into how CourseGPT facilitates precise, updated, context-aware information retrieval and generation, setting a new standard for interaction within intelligent systems. Fig. \ref{fig:course-gpt-workflow} illustrates the workflow of CourseGPT, depicting the integrated data extraction, processing, embedding, and knowledge base utilization. This section also depicts the iterative process of embedding user prompts to reference relevant material in the knowledge base to provide relevant RAG-LLM answers.

\subsection{Retrieval Augmented Generation}
RAG-based techniques are crucial for LLM advancement. While LLMs have demonstrated impressive results, there is room for improvement, especially in handling specialized knowledge tasks. RAG addresses this by enhancing LLMs with external, up-to-date information sources. RAGs overcome outdated knowledge and hallucination by combining the LLM's intrinsic knowledge with vast repositories from external databases \cite{ragSurvey}. RAGs retrieve relevant document chunks based on searches utilizing semantic similarity calculations, resulting in more accurate and reliable outputs \cite{ragSurvey}. RAGs are built on retrieval, generation, and augmentation techniques.

This work utilizes LLMs' advanced language understanding and generation capabilities as the generative component of the RAG framework to discover critical differences between different LLM models. This approach ensures continuous knowledge updates and domain-specific integration, making RAG a promising solution for domain-specific LLMs.

\subsection{Embeddings}
The effectiveness of RAG depends on the quality and relevance of the retrieved passages used as input to the generator. To ensure the retrieval of pertinent information from the ARA documentation, we utilize conversation chains to start prompting and answering and employ text embeddings to create a retrieval knowledge base for the LLMs.

Text embeddings are critical in enhancing textual data's semantic understanding and representation. Extracted text chunks from the pre-processed data are transformed into high-dimensional vector representations, facilitating more nuanced similarity assessments and improving the accuracy of passage retrieval using the pre-trained  gte-large-en-v1.5 text embeddings model from AliBaba-NLP \cite{embeddings}. These embeddings support a context length of up to 8192. These embeddings were trained through multi-stage contrastive learning. The first stage involves a preliminary Masked Language Modelling pre-training on shorter 800M text pair lengths. The second stage involves data resampling, which reduces the short text proportions and a continuation of the Masked Language Modelling pre-training. This model was chosen due to the consideration of its size and its performance on the Massive Text Embedding Benchmark (MTEB) leaderboard \cite{mteb}.

CourseGPT trains the RAG-LLM model by incorporating relevant passages from the course textbook, course slide decks, syllabus, schedule, and supplementary materials. This process involves utilizing conversation chains and text embeddings to process and generate responses. Integrating the RAG component with advanced data extraction techniques enhances CourseGPT's ability to provide contextually rich and accurate responses to user queries relevant to the retrieved knowledge base data. The top-$p$ section headers are searched for retrieval, and the entire content section is retrieved. Top-$p$ sampling chooses from the smallest possible set of words whose cumulative probability exceeds the probability $p$. The probability mass is then redistributed among this set of words. The size of the set of words can dynamically increase and decrease according to the next word's probability distribution. We choose the top-$p$,  $(p = 0.95)$, samples for our approach.

\subsection{Mistral LLMs}
LLMs are a robust computational model for performing various natural language processing tasks. These models learn statistical relationships from vast text data during their intensive training. LLMs excel at text generation, a form of generative AI. LLMs predict the next token or word given an input text, creating coherent and contextually relevant output. The largest and most capable LLMs are built using a decoder-only transformer architecture \cite{decoder}. These models are trained on billions of parameters.

Mistral-7b is a 7-billion-parameter language model developed by Mistral AI and is engineered for superior performance and efficiency, which leverages grouped-query attention for faster inference, coupled with sliding window attention, to effectively handle sequences of arbitrary length with a reduced inference cost. Mistral-7b achieves a 2x speed improvement for sequences up to 16k with a window of 4k due to its innovative sliding window attention mechanism \cite{mistral}.

Mixtral-8x7b is a Sparse Mixture of Experts language model developed by Mistral AI. Mixtral-8x7b has the same architecture as Mistral-7B, except each layer consists of 8 feedforward blocks. At each layer, a router network selects two feedforward blocks for every token to process the current state and combine their outputs. Each token can access 47B parameters but only uses 13B active parameters during inference due to feedforward blocks. Mixtral-8x7b utilizes a subset of its parameters for each token, allowing for faster inference speeds at low batch sizes and higher throughput at large batch sizes. Mixtral-8x7b's efficient parameter usage ensures competitive performance. Mixtral-8x7b was trained with a context size of 32k tokens \cite{mixtral}.

\subsection{Retrieval Material and Knowledge Base}
The training materials sourced for CourseGPT are essential to refining the RAG-LLM model. The retrieval material includes the course textbook, slide decks, the syllabus and schedule, and supplementary resources such as API documentation introductions to command modules. These materials constitute a rich knowledge base relevant to the course curriculum.

The training materials offer comprehensive course content coverage with details on topics, concepts, and methodologies relevant to the course. They include APIs associated with the course material, resource specifications, in-depth information on course-related topics, and supplementary information provided to students as optional readings.

The research materials were pre-processed and cleaned before fine-tuning. Our data extraction process involved parsing and organizing textual information into a cohesive, uniform format to enhance consistency and readability. Redundant or extraneous content, such as duplicate entries, formatting irregularities, and non-informative sections, were removed.

\section{Evaluation Methodology}
\label{methodology}

Establishing clear and measurable criteria to evaluate employed LLM efficacy within the course context is crucial. This section details assessment metrics for the LLMs' correctness scores, faithfulness, and context recall. Each metric provides valuable insights into the LLMs' capabilities, enabling a comprehensive analysis of their practical utility in a real-world education context. Below, we detail the methodologies for computing these metrics.

\subsection{Correctness Scores}
We evaluated answer correctness to determine the accuracy of each LLM's responses to queries. This metric involved comparing the generated answer to the ground truth and assessing the level of alignment between the two. A higher score indicates greater accuracy and a closer match to the ground truth. Answer correctness is based on semantic similarity and factual accuracy, combined using a weighted scheme to determine the overall correctness score \cite{ragas, giskard}.

Answer semantic similarity assesses the semantic resemblance between the generated answer, $\mathbf{G_A}$, and the ground truth, $\mathbf{G_t}$. This evaluation is based on the ground truth and the answer\cite{ragas, giskard}. A higher score signifies a higher alignment between the generated answer and the ground truth. Answer semantic similarity is calculated using the embeddings model to vectorize the ground truth answer and the generated answer, then computing the cosine similarity, $S_\text{cos}$, between the two vectors. The cosine similarity of the embedded, vectorized generated answer, $\vec{\mathbf{G_A}}$, and the embedded, vectorized ground truth, $\vec{\mathbf{G_t}}$, is given as follows:
\begin{equation}
S_\text{cos} =\cos (\theta)=\frac{\vec{\mathbf{G_t}} \cdot \vec{\mathbf{G_A}}}{\|\vec{\mathbf{G_t}}\|\|\vec{\mathbf{G_A}}\|}.
\end{equation}

Factual similarity quantifies the factual overlap between the generated answer and the ground truth answer as follows:
\begin{equation}
\text{F} = \frac{|\text{TP}|}{(|\text{TP}| + 0.5 \times (|\text{FP}| + |\text{FN}|))},
\end{equation}
where TP are true positives, facts, or statements present in both the ground truth and the generated answer, and FP are false positives, which are facts or statements present in the generated answer but not in the ground truth. FN are false negatives, facts, or statements present in the ground truth but not in the generated answer.

Answer correctness is given by the weighted average of $S_\text{cos}$ and F, where $W_{S_\text{cos}}=0.25$ and $W_\text{F}=0.75$, as follows: 
\begin{equation}
    \text{Answer Correctness} = \frac{W_{S_\text{cos}}*S_\text{cos} + W_{\text{F}} * F}{W_{S_\text{cos}} + W_{\text{F}}}.
\end{equation}

\subsection{Context Recall}
Context recall measures the RAG-LLM's context alignment extent of the retrieved context with the ground truth. The ground truth attributable to the relevant context is $\mathbf{G_{tc}}$. Context recall is calculated using the ground truth and the retrieved context. 

Each sentence in the ground truth answer is analyzed to determine if it can be attributed to the retrieved context. Ideally, all sentences in the ground truth answer are attributable to the retrieved context.

Context recall is calculated as follows:
\begin{equation}
    \text{Context Recall} = \frac{|\mathbf{G_{tc}}|}{\text{Number of Sentences in }\mathbf{G_t}}
\end{equation}

\subsection{Faithfulness}
Among the metrics evaluated on the LLMs, one important measure is model faithfulness. This metric assesses the accuracy of the generated answer in terms of factual consistency with the provided context. It is calculated based on the answer and the retrieved context \cite{ragas, giskard}. To be considered faithful, the generated answer must make claims that can be logically deduced from the given context. A set of claims from the generated answer is identified and compared with the given context to assess the faithfulness of the LLMs. The answer is deemed faithful if all the claims can be inferred from the context. Faithfulness is expressed in terms of $N_{Gc}$, the number of claims in the generated answer that can be inferred from the given context, and $N_C$, the total number of claims in the generated answer, as:
\begin{equation}
    \text{Faithfulness}=\frac{|N_{Gc}|}{|N_C|}.
\end{equation}

\section{Evaluation Results}
\label{results}
This work assesses and compares the performance of the RAG-LLMs – Mistral-7b and  Mixtral-8x7b – in answering student questions related to specific course-related material. The evaluation metrics included correctness scores, answer relevancy, and faithfulness scores. These metrics provide insight into the RAG-LLM's capabilities and suitability for course-related tasks. The tests were conducted using an LLM evaluator. A test set of $N = 50$ questions was generated, and the LLM answers were evaluated on their overall correctness, context recall, and faithfulness. The knowledge base consists of categorized topics extracted by the evaluation tool to assess where CourseGPT requires improvements and further fine-tuning.

\subsection{Correctness Scores}
Larger models, such as Mixtral-8x7b, have a higher capacity to represent and process complex domain-specific knowledge, including intricate concepts, terminologies, and relationships present in the input context and the shared knowledge base. This capacity enables them to generate more accurate and contextually relevant responses across domain-specific queries. Larger models benefit from increased exposure to diverse examples and data during training. This exposure allows them to generalize better and adapt more effectively to varying input contexts and query types within the domain-specific knowledge domain. This exposure also increases accuracy and reliability in generating correct responses across different scenarios and use cases.

\begin{figure}
    \centering
    \includegraphics[scale = 0.45]{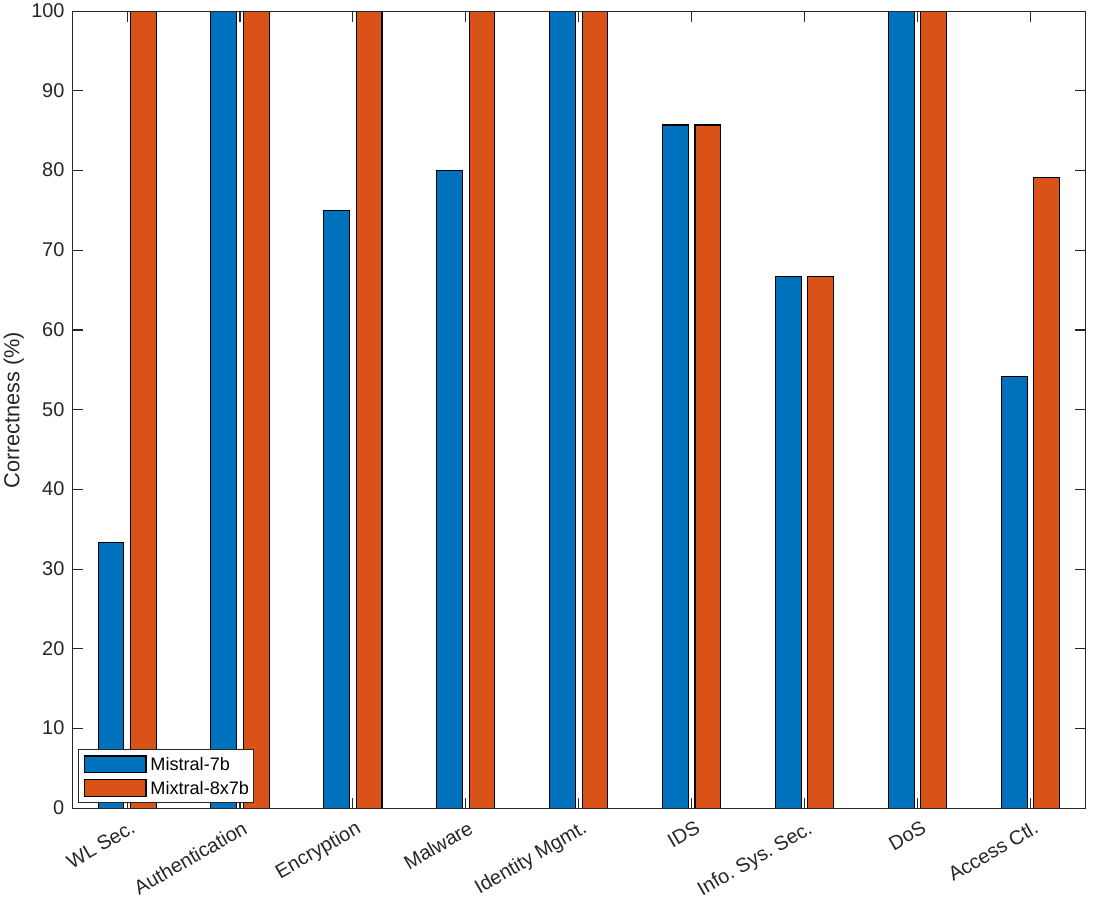}
    \caption{RAG-LLM correctness scores distribution}
    \label{correctness}
\end{figure}

Larger models exhibit greater robustness to noise, ambiguity, and variability in the input context and the shared knowledge base. They can filter irrelevant information and discern subtle differences, producing more precise and accurate responses.

The RAG-LLM model's correctness score distributions across different topics in our test set are shown in Fig. \ref{correctness}. The topics involved information regarding Wireless Security, Authentication, Encryption, Malware, Identity Management, Intrusion Detection Systems (IDS), Information System Security, Denial-of-Service (DoS), and Access Control, and other topics consisting of more general information not falling into any of the topics mentioned above. Fig. \ref{correctness} shows that the trend in higher correctness scores depends on the size of the model utilized with the same knowledge base. The "Others" topic results have been excluded from Fig. \ref{correctness} for simplicity.

\begin{figure}
    \centering
    \includegraphics[scale =0.45]{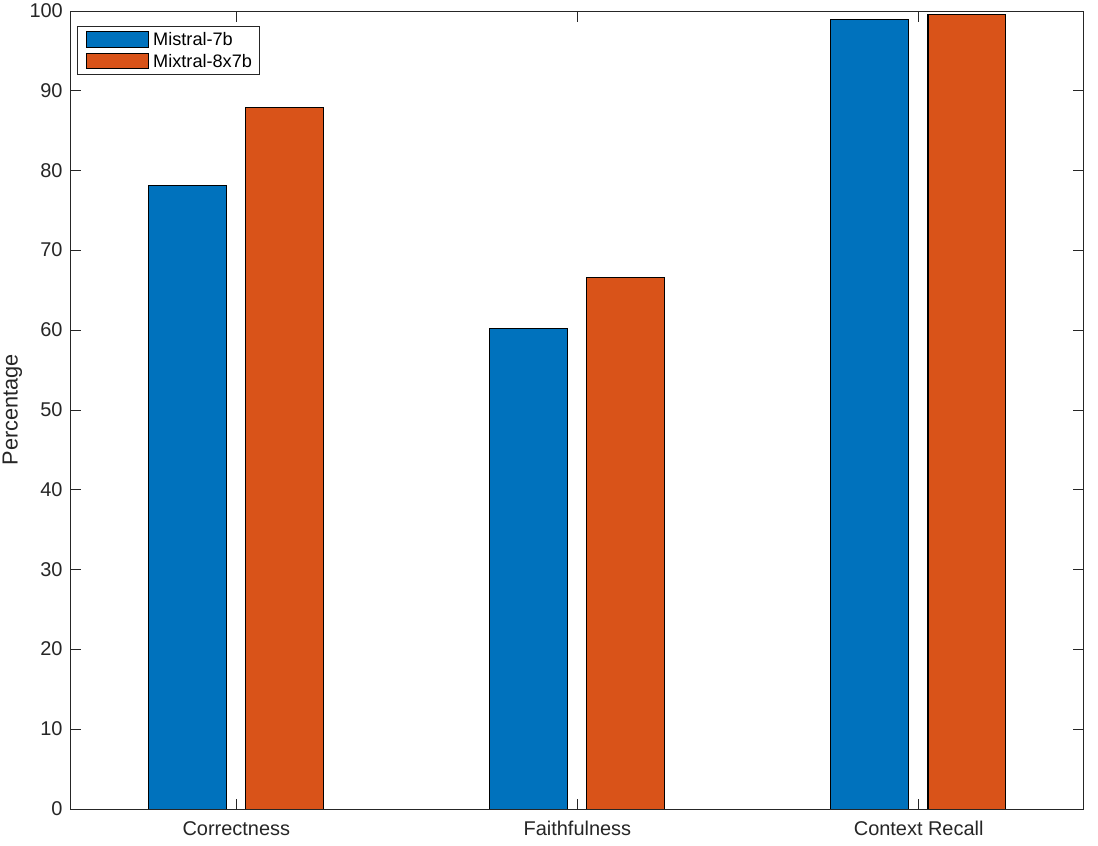}
    \caption{RAG-LLM performance metrics}
    \label{metrics}
\end{figure}

Our analysis revealed a significant difference in correctness scores between the RAG-LLMs. Mixtral-8x7b achieved the highest correctness score at 88.0\%, followed by Mistral-7b, which scored 78.2\%. This outcome shows that models with larger parameter sizes generally exhibit enhanced accuracy with higher correctness in generating contextually appropriate responses. It is worth noting that Mistral-7b performed as well as Mixtral-8x7b in the Authentication, Identity Management, IDS, Information System Security, and DoS topics despite being a smaller model, as seen in Fig. \ref{correctness}. The average RAG-LLM correctness scores for all topics are shown in Fig. \ref{metrics}.

The effectiveness of a model in generating correct responses is influenced by the distribution and training data coverage related to different topics within the knowledge base. Both models were exposed to diverse and representative examples for specific topics during training. As such, the smaller model learns to generalize and perform comparably to the larger model in those areas.

Smaller models may benefit from greater sample efficiency, requiring fewer examples or less training data to learn effectively. The smaller model may extract and leverage relevant patterns and information efficiently, leading to comparable performance to the larger model if the training data for specific topics is limited but informative.

The LLM model performance depends on its size and the optimization techniques and strategies employed during training. Effective regularization, fine-tuning, and optimization strategies can mitigate the limitations of smaller models and enhance their performance across various topics within the knowledge base. Both models share the same underlying architecture.

\subsection{Context Recall}
The RAG-LLMs demonstrated commendable context recall performance in our analysis, indicating a strong alignment with the source material and minimal deviation. However, minor performance differences between the models were notable due to the LLM's size. In the analysis, context recall was similar based on the answers given on the knowledge base. This finding suggests that models with higher parameter counts might be better at adhering strictly to the information provided in the knowledge base. However, the improvement is only slight since a shared knowledge base is used, and both LLMs perform considerably well while respecting the retrieved context.

Analysis revealed that, on average, Mixtral-8x7b was more adept at generating relevant answers across the tested questions than its counterpart, but only slightly. Fig. \ref{metrics} illustrates the distribution of context recall scores for the RAG-LLMs and shows how closely aligned the retrieved LLM context was to the ground truths. These results show that each RAG-LLM provided aligning and relevant information to answer prompts. Mixtral-8x7b scored an average context recall score of 99.6\%, while Mistral-7b scored 99.0\%. These findings show that the LLMs provide aligning and relevant information given a specific context. The context recall scores are similar due to the LLMs' underlying model architecture and the shared knowledge base usage.

\subsection{Faithfulness}
The RAG-LLM faithfulness evaluation showed that Mixtral-8x7b scored the highest in model faithfulness, followed by Mistral-7b. Mixtral-8x7b attained a faithfulness score of 66.6\%, and Mistral-7b scored 60.2\%. Our evaluation shows that using larger LLMs increases the faithfulness scores considerably and that the generated answer accuracy regarding factual consistency with the provided context is much higher than in smaller model sizes. A larger LLM has a higher capacity to capture and understand complex patterns in language due to its increased number of parameters and layers. This enhanced capacity enables larger LLMs to generate text that more closely resembles human-authored text and maintains higher faithfulness to the input prompt or context. 

In domain-specific applications like CourseGPT, accuracy and reliability are crucial. Students rely on CourseGPT for accurate information and guidance on course content, assignments, and exams. A larger LLM's higher faithfulness means it is more likely to provide accurate and relevant responses that align closely with the input provided by students or instructors. Faithfulness scores of the RAG-LLMs are shown in Fig. \ref{metrics}.

\subsection{Survey Results}
\begin{figure*}[!htb]
    \centering
    \begin{minipage}{2.25in}
        %\centering
        \centerline{\includegraphics[scale = 0.4]{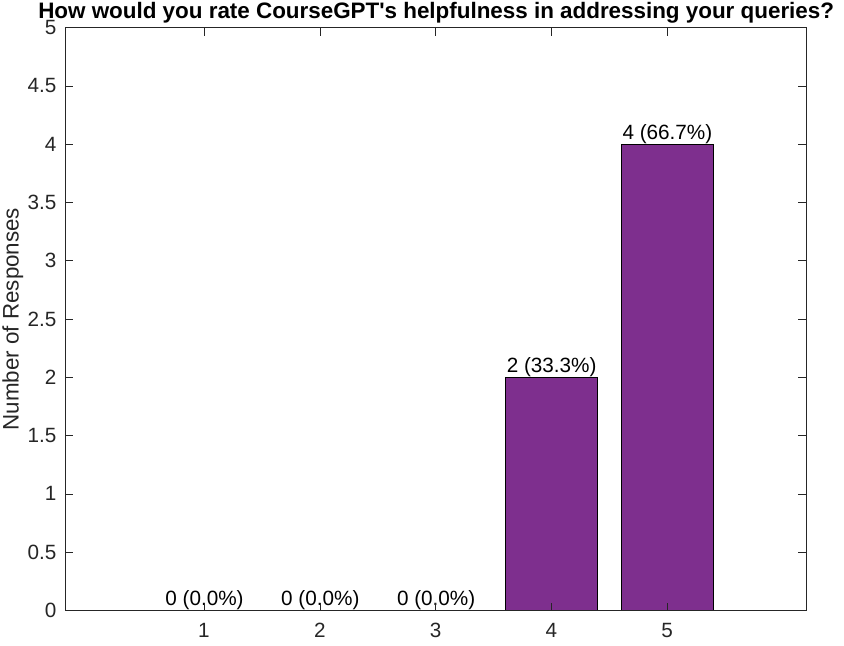}}
        \caption{Helpfulness}
        \label{fig:helpfulness}
    \end{minipage}
    \hspace*{0.1in}
    \begin{minipage}{2.25in}
        %\centering
        \centerline{\includegraphics[scale = 0.4]{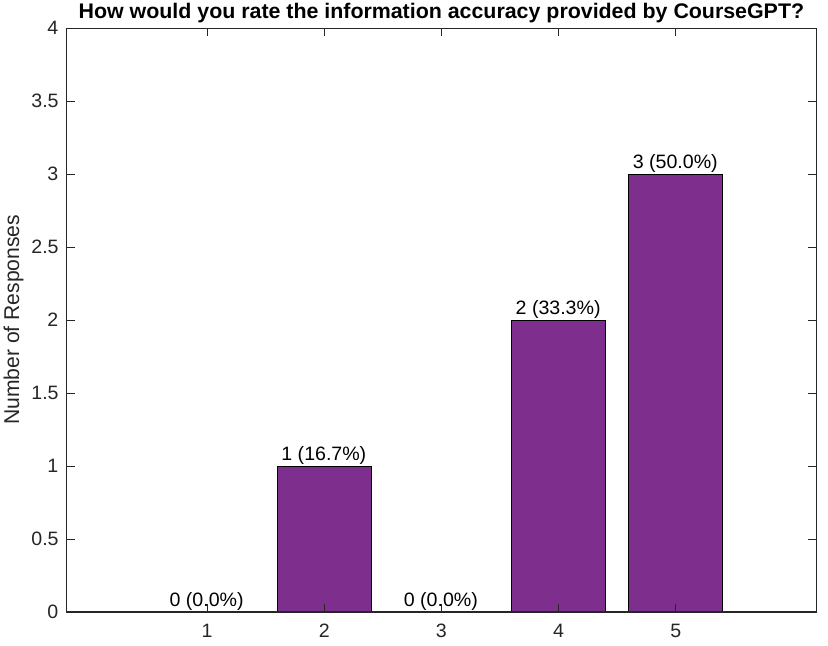}}
        \caption{Accuracy}
        \label{fig:accuracy}
    \end{minipage}
    \hspace*{0.1in}
    \begin{minipage}{2.25in}
        \centerline{\includegraphics[scale = 0.4]{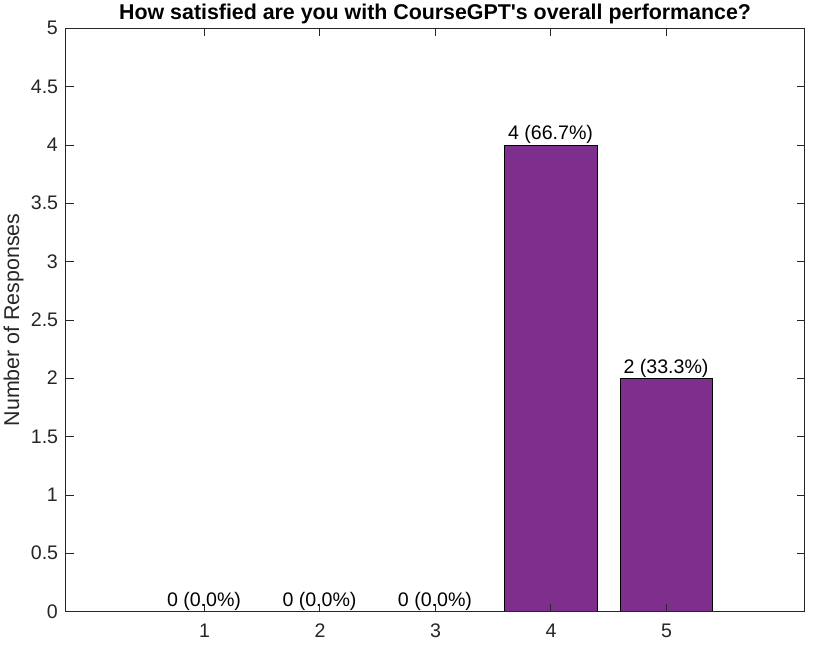}}
        \caption{Performance}
        \label{fig:performance}
    \end{minipage}
\end{figure*}

The survey assesses the satisfaction and performance of CourseGPT based on the feedback from six participants comprising former students and an experienced teaching assistant. The survey result analysis unveils valuable insights into CourseGPT's perceived effectiveness and utility in educational contexts. The survey questions were scored based on the 5-point Likert Scale. The survey consisted of 12 questions and were then combined into categories, as seen in the results in Figs. \ref{fig:helpfulness}, \ref{fig:accuracy}, \ref{fig:performance}, for simplicity.

The first survey question rated CourseGPT's helpfulness in addressing queries. Four participants rated CourseGPT with the highest score of 5, indicating exceptional helpfulness in addressing their queries. Two participants rated CourseGPT with a score of 4, affirming its commendable assistance in providing relevant and timely information. The first question's survey results are shown in Fig. \ref{fig:helpfulness}.

The second survey question asked for a rating of information accuracy provided by CourseGPT. The responses to this question demonstrate confidence in CourseGPT's ability to provide accurate information. Specifically, three participants rated CourseGPT with the highest score of 5, emphasizing their trust in the information accuracy delivered by the CourseGPT. Two participants rated CourseGPT with a score of 4, indicating solid reliability in the information provided. However, one participant rated CourseGPT with a score of 2, suggesting a need for improvement in ensuring information accuracy for specific queries. The second question's survey results are shown in Fig. \ref{fig:accuracy}.

The third survey question rated the participants' satisfaction with CourseGPT's overall performance. The survey findings indicate a mixed response regarding participants' satisfaction with CourseGPT's overall performance. While two participants rated CourseGPT with the lowest score of 2, suggesting areas for enhancement in performance, an equal number of participants rated CourseGPT with the highest score of 5, reflecting a high level of satisfaction with its overall performance. This rating divergence highlights the need for continuous refinement and optimization of CourseGPT to effectively address varying user expectations and preferences. The third question's survey results are shown in Fig. \ref{fig:performance}.

The survey results provide valuable insights into the perceived strengths and areas for improvement of CourseGPT in educational settings. Despite receiving high ratings for helpfulness and information accuracy from most participants, the mixed response regarding overall performance highlights the importance of ongoing refinement and enhancement efforts.

Addressing the concerns raised by participants who rated CourseGPT with lower scores will further enhance its effectiveness and user satisfaction. Additionally, leveraging the positive feedback and high ratings received in certain areas can serve as a foundation for promoting CourseGPT's adoption and utilization in future courses and educational endeavors.

A directed effort towards improving CourseGPT's performance, addressing user feedback, and refining its capabilities will be essential in solidifying its position as a valuable asset in modern education. CourseGPT can realize its potential as a transformative tool in streamlining personalized and compelling student and educator learning experiences by striving for optimal user experience.

\subsection{Result Discussions}
Analyzing different-sized LLMs offers valuable insights for optimizing workflows and improving retrieval support mechanisms for students enrolled in a course. The analysis highlights significant variances in correctness, context recall, and faithfulness scores among the LLMs, all determining CourseGPT's effectiveness in answering course-related queries.

The larger model, Mixtral-8x7b, which achieved higher correctness, context, and faithfulness scores, exemplifies superior proficiency in delivering precise and contextually relevant information. This precision is critical for students, enhancing their efficiency by enabling quicker access to necessary information and supporting informed decisions regarding their study materials and habits. 

\begin{figure}
    \centering
    \includegraphics[scale = 0.15]{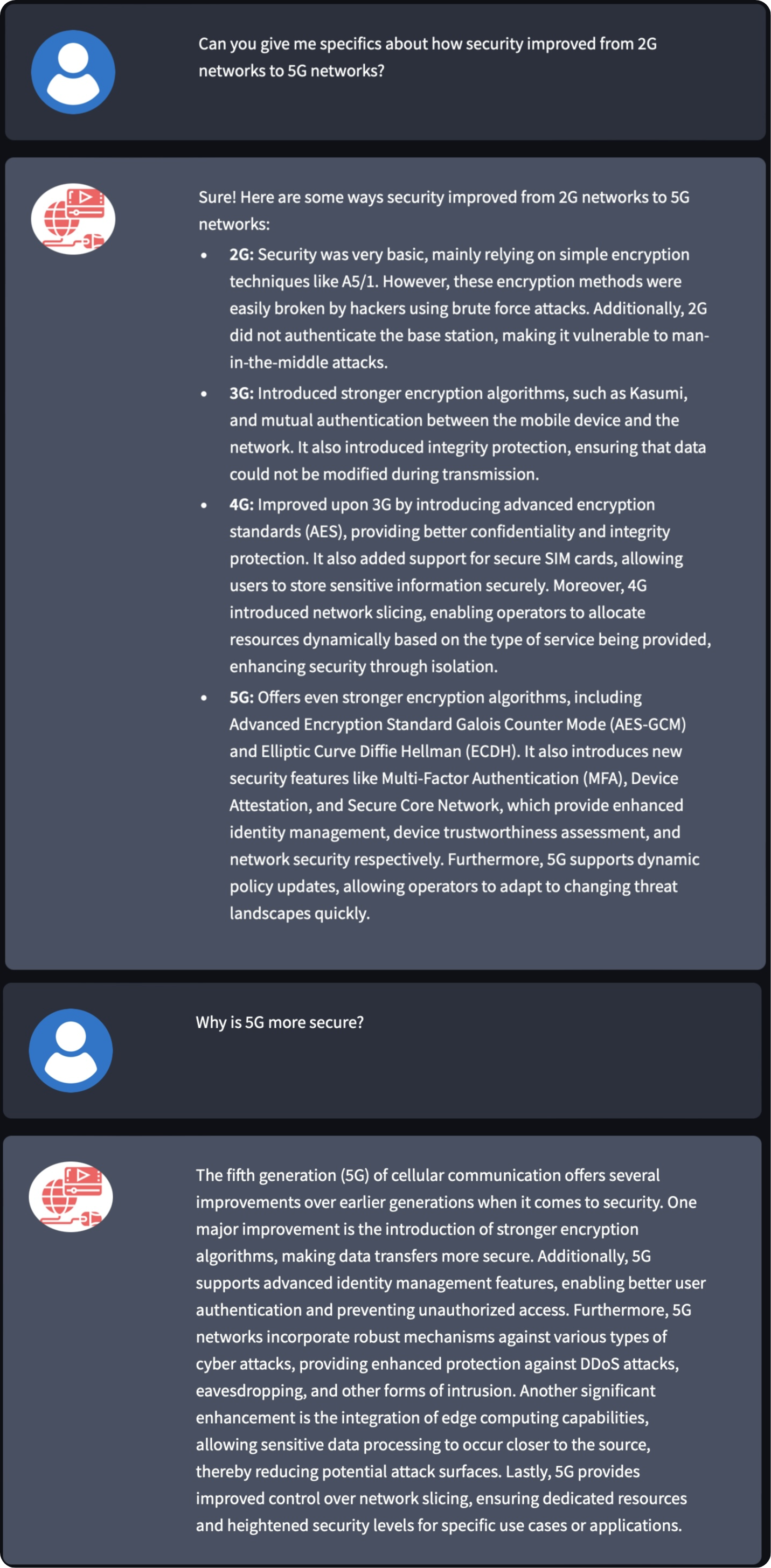}
    \caption{Initial and follow-up prompt to CourseGPT}
    \label{exampleoutput}
\end{figure}

These advantages enhance workflow efficiency and improve retrieval aids for students. By utilizing LLMs optimized for high correctness, context recall, and faithfulness to the source material, CourseGPT helps streamline workflow processes, and reduce the likelihood of errors or inaccuracies in answers. CourseGPT is a gateway for quickly accessing and retrieving course-related information, allowing students to gain more course-related information quicker. Fig. \ref{exampleoutput} shows an example output from the highest-overall performer, Mixtral8x7b.

%\begin{comment}
%\begin{figure}[h]
%    \begin{minipage}{\textwidth}
%        \includegraphics[scale=0.3]{Figures/prompt.png}
%        \caption{Initial and follow-up prompt to CourseGPT}
%        \label{exampleoutput}
%    \end{minipage}
%\end{figure}
%\end{comment}

% \begin{figure}[!htb]
%       \centering
%       % \begin{minipage}{6.9in}
%  \centerline{\includegraphics[scale = 0.4]{Figures/prompt.png}}
% \caption{Initial and follow-up prompt to CourseGPT}
%  \label{exampleoutput}
%       % \end{minipage}%
% \end{figure}

The implementation and testing of CourseGPT have yielded several significant implications for enhancing student learning outcomes and educational practices:

\subsubsection{Enhanced Instructor Efficiency}
Instructors benefit from the structured guidance CourseGPT offers, allowing them to focus more on high-level pedagogical strategies rather than routine question answering. This leads to more efficient use of class time and better allocation of instructional resources, ultimately improving the overall quality of education.

\subsubsection{Personalized Learning Experiences}
CourseGPT's ability to provide individualized feedback and tailored learning pathways supports diverse student needs, leading to more personalized learning experiences. This adaptability helps in addressing different learning paces and styles, contributing to more inclusive and effective education.

The results demonstrate that CourseGPT significantly enhances student engagement, performance, and personalized learning experiences while improving instructor efficiency and promoting critical thinking. These implications highlight the potential of CourseGPT to transform educational practices and contribute to higher standards of education.

\section{Challenges}
\label{challenges}
Integrating CourseGPT is a significant advancement in enhancing education. Several challenges require addressing to maximize the effectiveness of CourseGPT's goals. Identifying and mitigating these challenges allows CourseGPT to reach its full potential as an influential student assistant. This integration also results in enhanced productivity, efficient and quicker question-answering systems, and accelerated innovation.

\subsection{Memory Management and Computational Efficiency}
Deploying CourseGPT presents a challenge in memory management and computational efficiency. CourseGPT's reliance on extensive data and computational resources required to operate the LLMs and RAG effectively can cause memory issues, especially in environments with limited resources. The complex computations involved and the size of LLMs, such as those with more than 10 billion parameters, can strain system resources, resulting in latency issues, slowdowns, or system crashes. Memory management techniques and optimization strategies with supporting infrastructure can mitigate these challenges and ensure smooth operation.

\subsection{Scalability and Adaptability}
As the volume and complexity of student inquiries within the class scope grow, ensuring the scalability and adaptability of CourseGPT becomes essential. The system must be able to handle a diverse range of queries and research scenarios efficiently while maintaining responsiveness and accuracy.

Scalability challenges arise concerning LLM model sizes and the RAG framework's capacity to retrieve and generate relevant information in real-time. Addressing these challenges requires continuous optimization and refinement of CourseGPT architecture to accommodate evolving research requirements.

\subsection{Data Quality and Relevance}
A key obstacle we face is guaranteeing the quality and appropriateness of the data CourseGPT employs for knowledge retrieval and creation. The precision and scope of the class material, from which the data is directly extracted, directly affects CourseGPT's ability to offer valuable aid to students.

Sustaining data integrity and relevance in constantly evolving research environments, where new information is frequently generated and updated, presents persistent challenges. It is crucial to continue monitoring, updating, and validating course material and implement robust data preprocessing and filtering mechanisms to overcome these hurdles.

\section{Conclusion}
\label{conclusion}
%We generated responses using open-source LLMs, embeddings, and cloud infrastructure with CourseGPT. Our results show that larger LLMs achieve better results, especially regarding correctness and faithfulness. Our future work will further improve CourseGPT's user experience, performance and information accuracy. Our analysis shows larger LLMs can improve performance; however, infrastructure and resource constraints are worth considering, given the application and deployment of CourseGPT. With the advancements of AI in an educational environment that assists students and instructors, CourseGPT acts as a pillar advocating for carrying the next generation of education tools for a more personalized learning experience.
CourseGPT signifies a significant advancement in integrating generative AI into educational environments, demonstrating notable improvements in student engagement and learning outcomes. By leveraging the power of large language models, CourseGPT provides accurate, context-aware, and immediate responses to student inquiries, thus fostering a more interactive and personalized learning experience. Implementing CourseGPT in courses such as CPR E 431 has shown promising results, with increased correctness and faithfulness scores, underscoring the potential of large models like Mixtral-8x7b in enhancing educational support systems.
Moving forward, we will focus on refining CourseGPT’s capabilities and addressing challenges related to computational efficiency, scalability, and data relevance. Continuous improvements and adaptations will be necessary to ensure that CourseGPT remains a robust and reliable tool for instructors and students. As AI technology progresses, CourseGPT is poised to become an essential component of modern educational frameworks, promoting excellence in educational experiences and contributing to the broader goals of knowledge enhancement and discovery at every higher educational school.
Overall, CourseGPT exemplifies the transformative potential of AI in education, setting a precedent for future innovations that aim to create more dynamic, responsive, and effective learning environments.

\balance
%\newpage

\bibliographystyle{IEEEtran.bst}
\bibliography{bibliography.bib}

\end{document}